\documentclass[aps,prl,twocolumn,10pt,showpacs,superscriptaddress,amsmath,amssymb,nobibnotes]{revtex4-1}
\usepackage{graphicx}
\usepackage{amsfonts}    
\usepackage{graphicx}   
\usepackage{verbatim}   
\usepackage{color}      
\usepackage{subfigure}  
\usepackage{hyperref}   
\usepackage{natbib}
\usepackage{booktabs}
\usepackage{threeparttable}
\usepackage{dcolumn}
\usepackage{multirow}
\usepackage{float}
\usepackage{ulem}
\usepackage{bm}
\usepackage{algorithm}
\usepackage{algorithmicx}
\usepackage[noend]{algpseudocode}
\usepackage{CJK}

\makeatletter
\def\BState{\State\hskip-\ALG@thistlm}
\makeatother
\begin{document}
\begin{CJK*}{UTF8}{gbsn}


\title{Digital Holographic Imaging via Direct Quantum Wavefunction Reconstruction }

\author{Meng-Jun Hu (\CJKfamily{gbsn}胡孟军)}
\email{humj@baqis.ac.cn}
\affiliation{Beijing Academy of Quantum Information Sciences, Beijing 100193, China}

\author{Yong-Sheng Zhang (\CJKfamily{gbsn}张永生)}%
\email{yshzhang@ustc.edu.cn}
\affiliation{Laboratory of Quantum Information, University of Science and Technology of China, Hefei 230026, China }%
\affiliation{Synergetic Innovation Center of Quantum Information and Quantum Physics,
University of Science and Technology of China, Hefei 230026, China}
\affiliation{Hefei National Laboratory, University of Science and Technology of China, Hefei, 230088, China}

\date{\today}

\begin{abstract}
Wavefunction is a fundamental concept of quantum theory. Recent studies have shown surprisingly that wavefunction can be directly reconstructed via the measurement of weak value. The weak value based direct wavefunction reconstruction not only gives the operational meaning of wavefunction, but also provides the possibility of realizing holographic imaging with a totally new quantum approach. Here, we review the basic background knowledge of weak value based direct wavefunction reconstruction combined with recent experimental demonstrations. The main purpose of this work focuses on the idea of holographic imaging via direct wavefunction reconstruction. Since research on this topic is still in its early stage, we hope that this work can attract interest in the field of traditional holographic imaging. In addition, the wavefunction holographic imaging may find important applications in quantum information science.
\\

{\bf Keywords:} Wavefunction reconstruction, Weak value, Hologram imaging

{\bf PACS:} 03.67.-a, 42.50.Ex
\end{abstract}

\maketitle
\end{CJK*}


\section{1. Introduction}
Wavefunction or state vector is the core concept in quantum theory, which is believed to give complete description of a quantum system \cite{Dirac}. The probabilistic interpretation of wavefunction causes continuous arguments on its underlying physical meaning from the very beginning to now. On the one hand, the evolution of the wavefunction is totally determined by the Schr\"{o}dinger equation. On the other hand, there exists so called wavefunction collapse during the quantum measurement process. This duality has led people to ask whether wavefunction represents physical reality or is simply a reflection of our ignorance of the quantum system \cite{Beyasian1, Beyasian2, Beyasian3}. Since all interpretations are required to give the same experimental predictions, there seems to be no hope of settling down the argument in a short period of time. However, significant progress in the direct measurement of wavefunction has been made in the past decades due to the study of quantum weak measurements \cite{weak}.  The direct measurement of wavefunction experimentally gives a clear operational definition, which makes wavefunction from an abstract concept become a measurable object.  The meaning of direct wavefunction reconstruction is not limited to the fundamental aspect; its potential applications are also being explored, e.g., holographic imaging discussed in this work.

The possibility of direct wavefunction reconstruction is attributed to the study of quantum weak measurements and weak value, which was proposed by Aharonov, Albert and Vaidman (AAV) in 1988 \cite{AAV}. In 2011, the experimental realization of photonic one-dimensional spatial wavefunction was first demonstrated \cite{Ludeen} mainly due to the rapid development of quantum optics and quantum information.
In the standard quantum measurement process, the state $|\psi\rangle$ of a measured system will randomly collapse into one eigenstate $\lbrace|a_{k}\rangle\rbrace$ of observable $\hat{A}$ with probabilities according to the Born rule that $P_{k}=|\langle a_{k}|\psi\rangle|^{2}$. This projective measurement outputs only the amplitude of the wavefunction but not phase information. To tackle this issue, quantum state tomography (QST) may be used to reconstruct the quantum state via multiple basis measurements \cite{QST1, QST2, QST3}. Unfortunately, the method requires $o(n^{2})$ basis measurements for an $n$ dimensional quantum state, which makes wavefunction reconstruction time-consuming. In the framework of AAV weak measurements, however, the case of extremely weak coupling between the system and measurement apparatus is considered such that the system is almost undisturbed. In order to extract the information of system, post-selection of the system is introduced, and the weak value of observable $\hat{A}$ can be defined as $<\hat{A}>_{w}=\langle\psi_{f}|\hat{A}|\psi_{i}\rangle/\langle\psi_{f}|\psi_{i}\rangle$ with $|\psi_{i}\rangle, |\psi_{f}\rangle$ are the initial and post-selected states of the system, respectively. Different from the expectation value of an observable, the weak value can be complex and beyond the spectrum of eigenvalues. When $\langle\psi_{f}|\psi_{i}\rangle\to 0$, $<\hat{A}>_{w}$ can be very large, which makes weak value amplification be applied to various small signal detections \cite{am1, am2, am3}. More importantly, if we choose $\hat{A}=|x\rangle\langle x|$ and $|\psi_{f}\rangle=|p=0\rangle$, then $<\hat{A}>_{w}=\langle p=0|x\rangle\langle x|\psi_{i}\rangle/\langle p=0|\psi_{i}\rangle\propto \psi_{i}(x)$ with $\langle p=0|x\rangle=e^{-ip_{0}x}=1$ and $\langle p=0|\psi_{i}\rangle$ is a constant. Since the real and imaginary part of the weak value can be directly measured \cite{Jozsa}, the wavefunction $\psi_{i}(x)$ can thus be reconstructed from the measurement of the weak value. After the first experimental demonstration of direct wavefunction reconstruction, further research on this topic shows that strong coupling measurement with post-selection can also achieve the same goal \cite{strong, direct, zhang}. The weak value based wavefunction reconstruction is thus independent of measurement strength. 
With the ability of direct wavefunction reconstruction, digital holographic imaging of an object seems natural if we compare the photonic wavefunction $\psi_{o}(x, y)$ obtained after the object with the known input wavefunction $\psi_{i}(x, y)$. This is totally different from classical digital holograph imaging, in which the phase of the light field needs to be estimated by using interference \cite{holography1, holography2, holography3}. Although direct wavefunction reconstruction is proposed based on the theory of quantum measurements, it also applies to traditional light sources due to the bosonic nature of photons.  It is what we call the ``quantum-inspired" method, in which the essential idea is based on quantum research, but practical realization can also be realized only with classical instruments and technology. It provides different perspectives on the same problem from the point view of quantum measurement but is executable in practice with the existing technologies. 

In this work, we provide a basic introduction on the topic of direct reconstruction of wavefunction mainly based on our previous works, and discuss the opportunities and challenges in digital holographic imaging via direct wavefunction reconstruction. The paper is arranged as follows. In section II, we introduce the basic framework of weak measurements and the concept of weak value, which is the foundation of direct wavefunction reconstruction. Equipped with basic background knowledge, we introduce in section III the weak value based one-dimensional spatial wavefunction direct reconstruction first demonstrated in 2011. We then introduce the recent progress on two-dimensional spatial wavefunction direct reconstruction via strong measurements in section IV. In section V, we introduce the idea of scan-free wavefunction reconstruction to improve the speed of imaging. We give the general process to realize holographic imaging of objects based on direct wavefunction reconstruction in section VI. Challenges and opportunity are discussed in section VII.

\section{2. Weak Measurements and weak value}
In the standard textbook description of quantum mechanics, the system of state $|\psi_{s}\rangle$ will be collapsed into one of eigenstates $\lbrace|a_{k}\rangle\rbrace$ of observable $\hat{A}$ with probability determined by the Born rule $P_{k}=|\langle a_{k}|\psi_{s}\rangle|^{2}$. This kind of measurement, which is called projective or destructive measurement, is destructed because after measurement, the information encoded in the original state $|\psi_{s}\rangle$ has been totally destroyed. However, it provides most information about the system after measurement, i.e., the state of system is totally determined once we know the measurement outcome. In practice, however, projective measurement is only a special case of measurement because it requires the coupling between the pointer and the measured system be strong enough such that different eigenstates can be distinguished. There always exists measurement errors due to imperfections in the measurement apparatus. The general description of quantum measurement, which is more related to realistic physical realization, is given by the positive-operator valued measure (POVM) \cite{POVM}. The elements of POVM are $\lbrace \hat{E}_{k}\equiv\hat{M}^{\dagger}_{k}\hat{M}_{k}\rbrace$ satisfying completeness condition $\sum_{k}\hat{E}_{k}=I$, in which $\lbrace\hat{M}_{k}\rbrace$ are Kraus operators. Suppose the initial state of the system is $\rho_{s}$, then the probability of obtaining outcome $k$ is $P_{k}=\mathrm{Tr}(\hat{E}_{k}\rho_{s})$, and the corresponding state of the system after measurement becomes
\begin{equation}
\rho_{s}^{k}=\frac{\hat{M}_{k}\rho_{s}\hat{M}^{\dagger}_{k}}{\mathrm{Tr}(\hat{M}_{k}\rho_{s}\hat{M}^{\dagger}_{k})}=\frac{\hat{M}_{k}\rho_{s}\hat{M}^{\dagger}_{k}}{P_{k}}.
\end{equation}
Obviously, the projective measurement is included in the POVM if we take $\rho_{s}=|\psi_{s}\rangle\langle\psi_{s}|$ and $\hat{M}_{k}=|a_{k}\rangle\langle a_{k}|$.

\begin{figure}[tbp]
\centering
\includegraphics[scale=0.58]{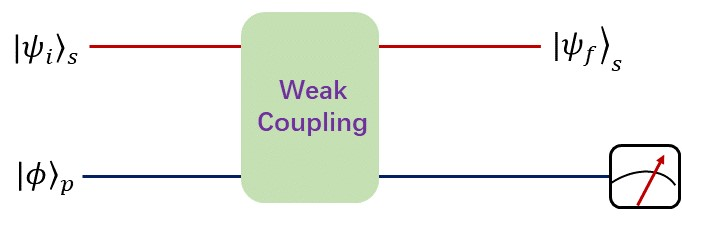}
\caption{Basic framework of weak measurements with post-selection. The coupling between the system and the pointer is weak enough such that weak value $<\hat{A}>_{w}=\langle\psi_{f}|\hat{A}|\psi_{i}\rangle_{s}/\langle\psi_{f}|\psi_{i}\rangle_{s}$ can be defined under the first order approximation. The suitable measurements on pointer, after post-selection of the system, extract the information of weak value. }
\label{f1}
\end{figure}

In 1988, Aharonov, Albert and Vaidman first introduced weak measurements with post-selection on the state of system, in which the coupling between the system and pointer is much smaller than the uncertainty of the system itself. The basic framework of weak measurements is shown in Fig. \ref{f1}. Suppose that the initial states of the system and the pointer are $|\psi_{i}\rangle_{s}$ and $|\phi_{i}\rangle_{p}$ respectively, the initial state of the composite system is $|\Psi_{i}\rangle_{sp}=|\psi_{i}\rangle_{s}\otimes|\phi_{i}\rangle_{p}$ given that the two systems are uncorrelated.
The interaction Hamiltonian between the system and the pointer is generally described as
\begin{equation}
\hat{H}_{sp} = g(t)\hat{A}\otimes\hat{P},
\end{equation}
where $g$ is the coupling coefficient, $\hat{A}$ is the observable to be measured, and $\hat{P}$ is the momentum operator of the pointer. After interaction, the composite system evolves as
\begin{equation}
|\Psi_{f}\rangle_{sp} = e^{-i\int g(t)\hat{A}\otimes\hat{P}dt}|\Psi_{i}\rangle_{sp}=e^{-ig_{0}\hat{A}\otimes\hat{P}}|\psi_{i}\rangle_{s}\otimes|\phi_{i}\rangle_{p},
\end{equation}
where pulse-type interaction $g(t)=g_{0}\delta(t-t_{0})$ with $g_{0}\ll 1$ is considered and $\hbar=1$ is used. We then consider the post-selection of the system of the final state $|\psi_{f}\rangle_{s}$, the state of the pointer will collapse into (unnormalized)
\begin{align}\nonumber
|\phi_{f}\rangle_{p} &=_{s}\langle\psi_{f}|e^{-ig_{0}\hat{A}\otimes\hat{P}}|\psi_{i}\rangle_{s}\otimes|\phi_{i}\rangle_{p} \\
&\approx _{s}\langle\psi_{f}|(1-ig_{0}\hat{A}\otimes\hat{P})|\psi_{i}\rangle_{s}\otimes|\phi_{i}\rangle_{p}\nonumber\\
&=_{s}\langle\psi_{f}(1-ig_{0}\frac{\langle\psi_{f}|\hat{A}|\psi_{i}\rangle_{s}}{\langle\psi_{f}|\psi_{s}\rangle_{s}})|\phi_{i}\rangle_{p}\nonumber \\
&\approx \langle\psi_{f}|\psi_{i}\rangle_{s}e^{-ig_{0}<\hat{A}>_{w}\hat{P}}|\phi_{i}\rangle_{p},
\end{align}
in which the first order approximation is used and the weak value of observable $\hat{A}$ is defined as
\begin{equation}
<\hat{A}>_{w}=\frac{\langle\psi_{f}|\hat{A}|\psi_{i}\rangle_{s}}{\langle\psi_{f}|\psi_{i}\rangle_{s}}.
\end{equation}
In the position representation, the state of the pointer becomes
\begin{equation}
\phi_{f}(x)=\langle x|\phi_{f}\rangle_{p}=\langle\psi_{f}|\psi_{i}\rangle_{s}\cdot\phi_{i}(x-g_{0}<\hat{A}>_{w}),
\end{equation}
which implies that the pointer has moved to a new position $g_{0}<\hat{A}>_{w}$. If the post-selection state $|\psi_{f}\rangle_{s}$ is chosen such that $\langle\psi_{f}|\psi_{i}\rangle_{s}\rightarrow 0$, then $<\hat{A}>_{w}\rightarrow\infty$ and $g_{0}<\hat{A}>_{w}$ would be sufficiently large. When the signal of interest is encoded in $g$, it can thus be significantly amplified by the weak value $<\hat{A}>_{w}$. However, it is at the cost of a low successful probability given by $|\langle\psi_{f}|\psi_{i}\rangle_{s}|^{2}$. The real and imaginary parts of the weak value $<\hat{A}>_{w}$ can be obtained by the measurement of suitable conjugate observables of the pointer \cite{Jozsa}.

In practical situations, the pointer is more suitable to be chosen as a qubit, e.g., polarization of photons, which is always the primary choice for photonic wavefunction reconstruction. The qubit pointer is usually spanned by the computational basis $\lbrace|0\rangle_{p}, |1\rangle_{p}\rbrace$ with the initial state prepared in $|0\rangle_{p}$. The interaction Hamiltonian should be replaced as $\hat{H}_{sp}=g(t)\hat{A}\otimes\hat{\sigma}_{y}$, where $\hat{\sigma}_{y}$ is the Pauli operator. The different eigenstates of observable $\hat{A}$ will cause the different rotation of the qubit in the Bloch sphere. The state of the qubit pointer, after post-selection of the system is recast as 
\begin{equation}
|\phi_{f}\rangle_{p}=\langle\psi_{f}|\psi_{i}\rangle_{s}\cdot e^{-ig_{0}<\hat{A}>_{w}\hat{\sigma}_{y}}|0\rangle_{p},
\end{equation}
which implies that the qubit is rotated with $2g_{0}<\hat{A}>_{w}$ in the Bloch sphere. The rotated state of the qubit has the form of $|\phi_{f}\rangle_{p}=\mathrm{cos}(\frac{\theta}{2})|0\rangle_{p}+e^{i\varphi}\mathrm{sin}(\frac{\theta}{2})|1\rangle_{p}$. It is thus physically clear to us that the angles $\theta, \varphi$ are determined by the real and imaginary parts of the weak value. Specifically, we have 
\begin{align}\nonumber
&\langle\phi_{f}|\hat{\sigma}_{+}|\phi_{f}\rangle_{p}=2g_{0}\mathrm{Re}<\hat{A}>_{w} \\
&\langle\phi_{f}|\hat{\sigma}_{R}|\phi_{f}\rangle_{p}=2g_{0}\mathrm{Im}<\hat{A}>_{w},
\end{align}
where $\hat{\sigma}_{+}\equiv|+\rangle\langle +|-|-\rangle\langle -|, \hat{\sigma}_{R}\equiv|R\rangle\langle R|-|L\rangle\langle L|$, and $|\pm\rangle=(|0\rangle\pm|1\rangle)/\sqrt{2}, |R/L\rangle=(|0\rangle\pm i|1\rangle)/\sqrt{2}$.

\section{3. Weak value based Wavefunction Reconstruction}

\begin{figure}[tbp]
\centering
\includegraphics[scale=0.38]{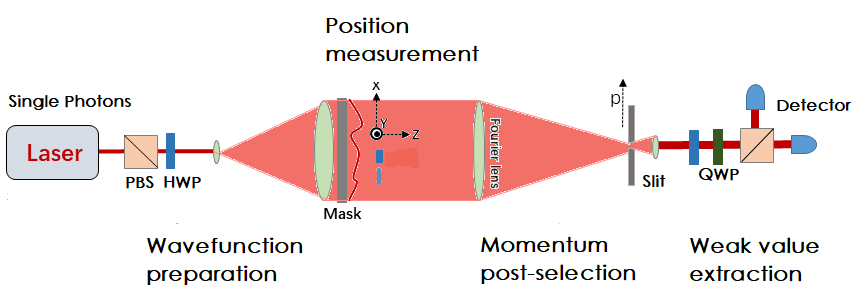}
\caption{Weak value based direct single photon one-dimensional spatial wavefunction $\Psi(x)$ reconstruction \cite{Ludeen}. }
\label{f2}
\end{figure}

The introduction of weak value and its strange structure provides the possibility of realizing direct wavefunction reconstruction. By letting the observable be position operator $|x\rangle\langle x|$ and choosing the post-selection state of the system as the zero momentum state $|p=0\rangle$, we obtain
\begin{equation}
<\hat{\pi}_{x}>_{w}=\frac{\langle p=0|x\rangle\langle x|\psi\rangle_{s}}{\langle p=0|\psi\rangle_{s}}=C\psi_{s}(x),
\end{equation}
where $\hat{\pi}_{x}\equiv|x\rangle\langle x|$ and $C=1/\phi_{s}(p=0)$ is a constant factor that can be renormalized. The real and imaginary parts of the wavefunction at position $x$ are directly determined by the real and imaginary parts of the weak value $<\hat{\pi}_{x}>_{w}$, respectively. The whole wavefunction $\psi_{i}(x)$ can be obtained by scanning the position. It is interesting to note that direct superposition wavefunction reconstruction is also possible with superposition position operator $\sum|x_{i}\rangle\langle x_{i}|$ \cite{superposition}. 

Although it seems direct to consider realization of wavefunction reconstruction from the definition of weak value, it was only experimentally demonstrated by J. S. Lundeen {\it et al} in 2011, which is already 23 years after the introduction of weak value. On the one hand, a new concept needs time to be widely accepted by physical communities; on the other hand, the rapid development of quantum information science requires us to have the ability to obtain more information about the quantum state. The widely adapted method of state reconstruction is quantum state tomography, which is indirect and requires multi-basis measurements to estimate the true quantum state. For an $n$ dimensional quantum state, at least $o(n^{2})$ different basis measurements are needed, which make it time-consuming for many qubits system. The direct wavefunction construction based on the weak value introduced here provides a promising way to tackle this difficulty.

The experimental setup to demonstrate the weak value based direct single photons spatial wavefunction reconstruction is shown in Fig. \ref{f2}. It consists of four parts, i.e., preparation of initial wavefunction, weak measurement of position $x1$, post-selection of system to zero momentum state, and readout of weak value. In the weak measurement, the polarization of photons is chosen as the pointer that is initially prepared in horizontal polarization state $|H\rangle$.  Weak measurement of position $x$ is performed by a rectangular small sliver of a half-wave plate (HWP) with size $x\times y\times z$ dimensions of $1\mathrm{mm}\times 25\mathrm{mm}\times 1\mathrm{mm}$. The weak coupling is realized by rotating the HWP with a small angle $\theta=20^{\circ}$. If $\theta=90^{\circ}$, the horizontal polarization of photons is transformed into vertical polarization, and we can definitely know which position the photons has passed through by its polarization information. The $\theta=90^{\circ}$ case thus corresponds to strong measurements or projective measurement of $|x\rangle\langle x|$. The interaction Hamiltonian is given as
\begin{equation}
\hat{H}_{sp}=\theta\hat{\pi}_{x}\otimes\hat{\sigma}_{y}.   
\end{equation}
The post-selection of the system to the zero momentum state is completed by a Fourier transformation lens placed on one focal length from the HWP and only are selected the photons with $p=0$ in the Fourier transform plane. According to Eq. (7), after the post-selection the polarization state of photons becomes
\begin{equation}
|\phi\rangle_{p}=\frac{1}{C}\cdot e^{-i\theta<\hat{\pi}_{x}>_{w}\hat{\sigma}_{y}}|H\rangle_{p}.
\end{equation}
The weak value $<\hat{\pi}_{x}>_{w}$ is obtained in the last stage by using a polarization analysis with direct connection to Pauli observables as
\begin{equation}
<\hat{\pi}_{x}>_{w}=\frac{1}{2\theta}[\langle\phi|\hat{\sigma}_{+}|\phi\rangle_{p}+i\langle\phi|\hat{\sigma}_{R}|\phi\rangle_{p}].
\end{equation}
For each position $x$, we need to perform only two projective measurements, and the full wavefunction $\psi(x)$ is directly reconstructed by scanning the position along the $x$ direction. Since the successful probability of post-selection is given by $1/C^{2}$, the linear scaling in the projective measurement is actually at the cost of more photons being measured. It is, however, not the problem in practice. 

In practice, the two-dimensional spatial wavefunction $\psi(x, y)$ is the most interesting, and can be used to realize holographic imagining of three-dimensional objects . The first experiment does not demonstrate that is because the $y$ dimension of the rectangular small sliver has to be used to move the sliver along the $x$ direction. It is thus vital to use an alternative setup to efficiently perform two-dimensional position $(x, y)$ measurements. The spatial light modulator (SLM) is a better choice to try.
In addition, the small angle $\theta$ rotated by the HWP has its own uncertainty which limits the accuracy of the weak value measurement. 
It is shown in the following that strong measurements with post-selection can do the same thing, and it provides more accurate results in principle. 

\begin{figure}[tbp]
\centering
\includegraphics[scale=0.58]{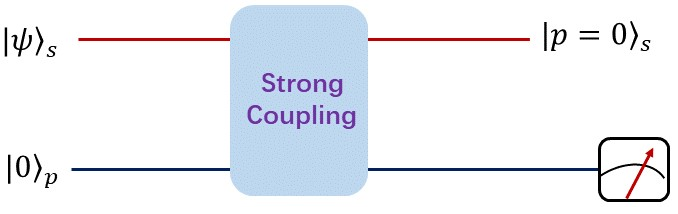}
\caption{Schematic diagram of direct wavefunction reconstruction via strong measurements with post-selection. }
\label{f3}
\end{figure} 

\begin{figure}[bp]
\centering
\includegraphics[scale=0.58]{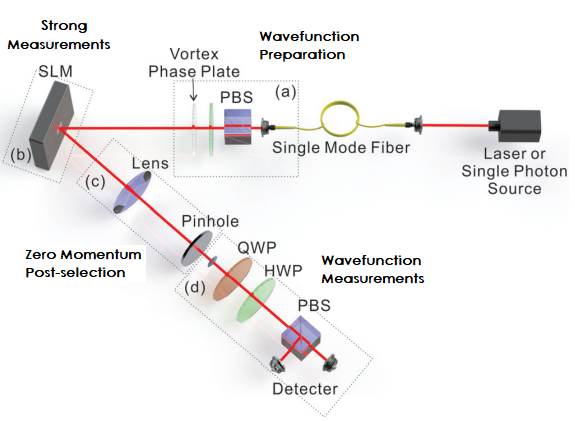}
\caption{Experimental setup for direct two-dimensional wavefunction $\psi_{s}(x, y)$ reconstruction via strong measurements and post-selection \cite{direct}. }
\label{f4}
\end{figure} 

\section{4. Two-dimensional Wavefunction Reconstruction via strong measurements}
We have introduced weak measurements and the weak value based direct spatial wavefunction reconstruction. In the previous experiments, weak value that related to wavefunction appears only given that coupling between the system and the pointer is weak enough. It causes two problems. First, the weak coupling in the actual experiment requires precision control of measurement apparatus, e.g., SLM; second, the weak value is obtained in the first order approximation, which brings naturally the uncertainty in the reconstruction of wavefunction.  In order to tackle this issues, people starts to think if strong measurements can complete the same target. 
G. Vallone and D. Dequal first consider the use of strong measurements with post-selection to realize direct wavefunction reconstruction\cite{strong}. In the original paper, they believe that weak value is not necessary within the strong measurement framework. However, it is not true. It only proves that weak value is independent of the coupling strength. We will show below how the weak value naturally appears in the strong measurements framework with post-selection to realize two-dimensional wavefunction reconstruction.

The basic framework, as shown in Fig. \ref{f3}, is almost the same as Fig. \ref{f1} but with weak coupling replaced with strong coupling. 
The pointer is chosen as a qubit and initially prepared in state $|0\rangle_{P}$. The interaction Hamiltonian is described by Eq. (10) but with $\theta=\pi/2$ for strong measurement. The composite system, after strong coupling interaction, becomes
\begin{align}\nonumber
|\Psi\rangle_{sp}&=e^{-i\frac{\pi}{2}\hat{\pi}_{x}\otimes\hat{\sigma}_{y}}|\psi\rangle_{s}\otimes|0\rangle_{p} \\
&=|\psi\rangle_{s}\otimes|0\rangle_{p}-\sqrt{2}\psi_{s}(x)|x\rangle\otimes|-\rangle_{p},
\end{align}
where $e^{-i\frac{\pi}{2}\hat{\pi}_{x}\otimes\hat{\sigma}_{y}}=\hat{I}-|x\rangle\langle x|\otimes(\hat{I}+i\hat{\sigma}_{y})$ is used. The post-selection of zero momentum state $|p=0\rangle=\sum_{x=1}^{d}|x\rangle/\sqrt{d}$ collapses the state of the pointer into
\begin{equation}
\begin{split}
|\phi\rangle_{p}&=\langle p=0|\Psi\rangle_{sp} \\
&=\phi_{s}(p=0)|0\rangle_{p}-\sqrt{2}\psi_{s}(x)\langle p=0|x\rangle|-\rangle_{p}  \\
&=\frac{1}{C}[|0\rangle_{p}-\sqrt{2}<\hat{\pi}_{x}>_{w}|-\rangle_{p}]
\end{split}
\end{equation}
The real and imaginary parts of weak value $<\hat{\pi}_{x}>_{w}$ can be extracted by suitable projective measurements on the pointer. Specifically, we have
\begin{align}\nonumber
&\mathrm{Re}[<\hat{\pi}_{x}>_{w}]=\frac{1}{2|C|^{2}}\langle\phi|(\hat{\sigma}_{x}+2|1\rangle\langle 1|)|\phi\rangle_{p} \\
&\mathrm{Im}[<\hat{\pi}_{x}>_{w}]=\frac{1}{2|C|^{2}}\langle\phi|\hat{\sigma}_{y}|\phi\rangle_{p},
\end{align}
where $\hat{\sigma}_{x}, \hat{\sigma}_{y}$ are Pauli operators defined in the $\lbrace|0\rangle, |1\rangle\rbrace$ basis.
Compared to the case of weak measurements, only one extra measurement, i.e., $|1\rangle\langle 1|$ is needed. The above derivations are all exact without any approximation, while the weak value is only naturally appears in the case of weak measurements under the first order approximation. The above analysis is totally applicable to two-dimensional position measurement with only $|x\rangle\langle x|$ replaced by $|x, y\rangle\langle x, y|$.

The first experimental realization of two-dimensional wavefunction $\psi_{s}(x, y)$ direct reconstruction of photons was reported by the group including current authors \cite{direct}, and the experimental setup is shown in Fig. \ref{f4}.  It consists of four parts, i.e., preparation of the spatial wavefunction of photons, strong measurements of position $(x, y)$, post-selection of photons to the zero momentum state and polarization measurement for extracting wavefunction information. The key improvement in our setup is the use of phase-type SLM to realize strong measurements of $|x, y\rangle\langle x, y|$. The initial state of the pointer is chosen as linear polarization state $|0\rangle_{p}=(|H\rangle+|V\rangle)/\sqrt{2}$. The phase-only SLM (Hamamatsu X10468) can convert the pointer state from $|0\rangle_{p}$ to orthogonal state $|1\rangle_{p}=(|H\rangle-|V\rangle)/\sqrt{2}$ by introducing $\pi$ phase difference between polarization states $|H\rangle$ and $|V\rangle$ from the reflection. In our experiment, the phase added area is $10\times 10$ pixels in the SLM with each pixel is about $12.5\mu m\times 12.5\mu m$. This completes the strong measurement interaction Hamiltonian $\hat{H}_{sp}=\theta|x, y\rangle\langle x, y|\otimes\hat{\sigma}_{y}$ with $\theta=\pi/2$ and the corresponding unitary $\hat{U}=e^{-i\hat{H}_{sp}}$. By moving the SLM step by step, the two-dimensional wavefunction in the imagine plane can be obtained. It should be noted that the coupling strength depends on the added phase and we need to keep it stable during the whole experiment. 
The post-selection of zero momentum state is realized by a Fourier lens with a pinhole of $50\mu m$ diameter placed in the focal place of the lens.
The suitable Pauli basis measurements with polarization analyzer, which consists of HWP, QWP and PBS, give us the information about $\psi_{s}(x, y)$ according to Eq. (15). It should be noted that $\hat{\sigma}_{x}, \hat{\sigma}_{y}$ are defined in the basis $\lbrace|0\rangle_{p}, |1\rangle_{p}$ with $|0\rangle(|1\rangle)=(|H\rangle\pm|V\rangle)/\sqrt{2}$ in the experiment. Full distribution of the wavefunction $\psi_{s}(x, y)$ is obtained by scanning along the $x-y$ plane of the SLM, which is mounted on a movable platform. 
We reported the direct measurement of spatial wavefunction for photons with angular momentum $l=1$, and the fidelity between the measured wavefunction and ideal wavefunction reaches approximately $0.93$. For more experimental details, please refer to Ref. \cite{direct}. 

\begin{figure}[tbp]
\centering
\includegraphics[scale=0.65]{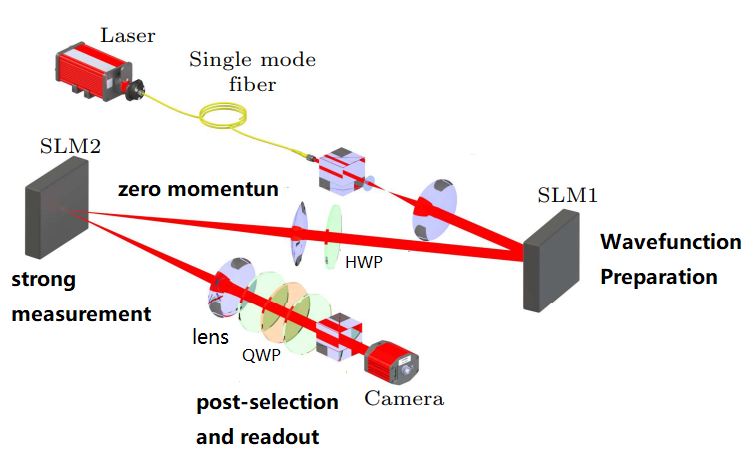}
\caption{Experimental setup of scan-free direct two-dimensional wavefunction $\psi_{s}(x, y)$ reconstruction via strong measurements \cite{zhang}. }
\label{f5}
\end{figure} 

\section{5. Scan-free wavefunction reconstruction}
In the previous wavefunction reconstruction, we need to scan the measurement plane step by step, which is time-consuming. This issue will limits our practical potential because it only applies to static or very slowly changing objects imaging. To address this issue, the scan-free method was proposed \cite{free}. The key idea is to change the role of position and momentum, i.e, perform the weak measurement of zero momentum and then execute position post-selection. In this case, we actually want to obtain the weak value of zero momentum
\begin{equation}
<\hat{\pi}_{p_{0}}>_{w}=\frac{\langle x|p=0\rangle\langle p=0|\psi\rangle_{s}}{\langle x|\psi\rangle_{s}}=\frac{1}{C\psi_{s}(x)},
\end{equation}
where $\hat{\pi}_{p_{0}}\equiv|p=0\rangle\langle p=0|$ and $C$ is the same as given in Eq. (9). Compared with the weak value of $<\hat{\pi}_{x}>_{w}$, it is easy to conclude that for any position and momentum satisfying the relation
\begin{equation}
<\hat{\pi}_{x}>_{w}\cdot<\hat{\pi}_{p}>_{w}=1.
\end{equation}
Since all the position measurements in the image plane can be realized by using detector arrays, there is no need for time-consuming scanning. The scan-free method makes dynamic object imaging practically possible.

We now examine whether this scan-free method works in the case of strong measurements, in which the interaction Hamiltonian becomes $\hat{H}_{sp}=\theta\hat{\pi}_{p_{0}}\otimes\hat{\sigma}_{y}$ with $\theta=\pi/2$. The composite state of the system and the pointer, after interaction, becomes
\begin{align}\nonumber
|\Psi\rangle_{sp}&=e^{-i\frac{\pi}{2}|p=0\rangle\langle p=0|\otimes\hat{\sigma}_{y}}|\psi\rangle_{s}\otimes|0\rangle_{p} \\
&=|\psi\rangle_{s}\otimes|0\rangle_{p}-\sqrt{2}\phi_{s}(p=0)|p=0\rangle\otimes|-\rangle_{p}.
\end{align}
The post-selection of the system with position $x$ collapses the pointer state into (unnormalized)
\begin{equation}
\begin{split}
|\varphi\rangle_{p} &= \langle x|\Psi\rangle_{sp}=\psi_{s}(x)|0\rangle_{p}-\sqrt{2}\phi_{s}(p=0)\langle x|p=0\rangle|-\rangle_{p}  \\
&=\psi_{s}(x)[|0\rangle_{p}-\sqrt{2}<\hat{\pi}_{p_{0}}>_{w}|-\rangle_{p}].
\end{split}
\end{equation}
The suitable basis measurements on the pointer gives
\begin{equation}
\begin{split}
\mathrm{Re}(<\hat{\pi}_{p_{0}}>_{w}) &=\frac{|\psi_{s}(x)|^{2}}{2}\langle\varphi|(\hat{\sigma}_{x}+2|1\rangle\langle 1|)|\varphi\rangle_{p}    \\
\mathrm{Im}(<\hat{\pi}_{p_{0}}>_{w}) &=\frac{|\psi_{s}(x)|^{2}}{2}\langle\varphi|\hat{\sigma}_{y}|\varphi\rangle_{p},
\end{split}
\end{equation}
The wavefunction $\psi_{s}(x)$ is thus obtained according to Eq. (16) with the constants be normalized. The above analysis is also applied to two-dimensional wavefucntion $\psi_{s}(x, y)$.

The experiment setup to demonstrate this scan-free direct wavefunction $\psi_{s}(x, y)$ reconstruction is shown in Fig. \ref{f5}. The setup consists of four parts, i.e., wavefunction preparation, strong interaction measurements of the zero momentum state, post-selection of position and polarization measurement to extract the weak value $<\hat{\pi}_{p_{0}}>_{w}$. In the experiment, we prepare photons with certain orbital angular momentum  using SLM1, and its polarization is prepared as $|0\rangle_{p}=(|H\rangle+|V\rangle)/\sqrt{2}$ with an HWP. Another phase-only SLM2 is placed on the focal length of the Fourier lens and converts the polarization only with zero momentum state into $|1\rangle_{p}=(|H\rangle-|V\rangle)/\sqrt{2}$, which completes the strong measurements of $\hat{\pi}_{p_{0}}$. 
Post-selection of positions is completed by placing a CMOS camera at the imagine plane of the $4f$ system. Photons that be recorded and counted in each pixel of camera, which corresponds to position measurement of photons determined by pixel position. Since each pixel works independently, post-selection of all positions in the imagine plane actually be completed in parallel, i.e., scan-free measurement.
At last a polarization analyzer is placed before the camera to realize polarization measurements to extract the momentum weak value. More experimental details can be seen in Ref. \cite{zhang}.

\begin{figure}[tbp]
\centering
\includegraphics[scale=0.65]{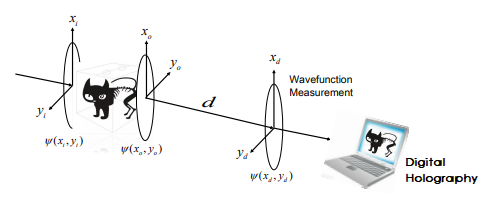}
\caption{Schematic diagram of digital holographic imaging based on wavefunction reconstruction. }
\label{f6}
\end{figure} 

\section{6. Holographic imaging via direct wavefunction reconstruction}

The possibility of direct two-dimensional wavefunction $\psi(x, y)$ reconstruction based on weak value naturally prompts us to consider its use in the holographic imaging of 3D objects. 
In conventional digital holographic imaging, the amplitude and phase information of object plane wavefront is usually obtained by introducing reference wave and let it interferes with object wave in the hologram plane. Suppose that the object and reference wave are described by complex amplitude field $O(x, y)$ and $R(x, y)$ respectively, the light intensity in the hologram plane is given by
\begin{equation}
\begin{split}
&H(x, y)=|O(x, y)+R(x, y)|^{2} \\
=&|O(x,y)|^{2}+|R(x,y)|^{2}+O(x,y)R^{*}(x,y)+O^{*}(x,y)R(x,y),
\end{split}
\end{equation}
where $O^{*}(x, y), R^{*}(x,y)$ are complex conjugate of $O(x, y), R(x, y)$. The first two parts are background terms, while the rest parts are interference terms. The object information is encoded in the third term $O(x,y)R^{*}(x,y)$ and the last term is its conjugate. In practice, we want to eliminate the background and conjugate terms, which can be done by using phase-shifted interferometry method with Mach-Zehnder or Michelson architecture \cite{phase}.

Compared with traditional holographic imaging, the quantum approach does not need the additional reference wave to interfere with object wave. The quantum approach takes advantage of quantum measurement framework with post-selection to realize the direct reconstruction of object wave $O(x,y)$. In quantum language $O(x, y)$ should be replaced with object wavefunction $\psi(x, y)$, but it makes no essential difference due to the bosonic nature of photons. Quantum interference is the key in the process of post-selection, which makes wavefunction holographic imaging totally different with traditional ways. 

Once we obtain the object wavefunction in the hologram plane, the next thing is to infer the photonic wavefunction in the object plane. 
Consider the imaging of a cat, as shown in Fig. \ref{f6}, with the input wavefunction $\psi(x_{i}, y_{i})$ of photons known, e.g., Gaussian distribution. The cat will change the wavefunction distribution of photons, and its full information is encoded in the amplitude and phase of object plane wavefunction $\psi(x_{o}, y_{o})$.
In the experiment, what we measure is the imagine plane wavefunction $\psi(x_{d}, y_{d})$, which has a distance $d$ from the object plane. The imagine plane wavefunction $\psi(x_{d}, y_{d})$ is determined by the object plane wavefunction $\psi(x_{o}, y_{o})$ via free-space Feynman propagator formula 
\begin{equation}
\psi(x_{d}, y_{d})=\int K(x_{d}, y_{d}, x_{o}, y_{o})\psi(x_{o}, y_{o})dx_{o}dy_{o},
\end{equation}
where the propagator is given by
\begin{equation}
K(x_{d}, y_{d}, x_{o}, y_{o})=\frac{1}{i\lambda}\frac{e^{ik\sqrt{(x_{d}-x_{o})^{2}+(y_{d}-y_{o})^{2}+d^{2}}}}{\sqrt{(x_{d}-x_{o})^{2}+(y_{d}-y_{o})^{2}+d^{2}}}
\end{equation}
with $\lambda, k$ are the wavelength and wavevector of photons respectively.
When $d\gg 1$, the paraxial condition is satisfied, and the $\psi(x_{o}, y_{o})$ can be obtained by inverse transformation
\begin{equation}
\psi(x_{o}, y_{o}) = \int L(x_{o}, y_{o}, x_{d}, y_{d})\psi(x_{d}, y_{d})dx_{d}dy_{d}
\end{equation}
with 
\begin{equation}
L(x_{o}, y_{o}, x_{d}, y_{d})=\frac{e^{-ikd}}{-i\lambda d}\mathrm{exp}\lbrace\frac{-ik[(x_{d}-x_{o})^{2}+(y_{d}-y_{o})^{2}]}{2d} \rbrace.
\end{equation}
By using the fast Fourier transformation algorithm, we can compute the object plane wavefunction $\psi(x_{o}, y_{o})$ from the measured imagine plane wavefunction $\psi(x_{d}, y_{d})$. 
Once the information of $\psi(x_{o}, y_{o})$ is obtained, the 3D reconstruction of the object is possible if the appropriate scattering model, or forward model is adapted. Specifically, it requires to map 3D refractive index distribution $n(x, y, z)$ in object space to the scattered wavefunction $\psi(x_{o}, y_{o})$. The forward model chosen is much like that in traditional ways, e.g., beam propagation model \cite{BPM}, the only difference is the replacement of complex light field with wavefunction in quantum approach. 
It should be emphasized here again that the above theoretical analysis is not limited to the quantum measurement case but also applies to conventional light sources due to the bosonic nature of photons.
There seem to be no practical difficulties in performing the above procedure with current technologies and algorithms in principle.

\section{7. Challenges and Opportunity}
Direct wavefunction reconstruction and its potential in digital holographic imaging are currently in the early stage of exploration. The theoretical framework has been completed, but the experimental realization still faces some technical difficulties. Although the imaging of light fields has already been demonstrated, holographic imagining of real objects requires more effort and technical improvements. Based on the previous experiments on light field imaging, there are still three main technical challenges to realize holographic imaging of 3D objects. The first is the coupling between the system and the pointer performed by the SLM. The extinction ratio of each pixel is actually different from each other, and we have to search the whole reflection area to find the best subarea of pixels to guarantee good coupling. This is not an easy thing in the practical experiment and lots of time has to be taken. The second is the post-selection of the zero momentum state or measurement of the zero momentum in the scan-free framework. In practice, we need to guarantee that there are no or few other momentum states coupled to the pointer.
The third is to develop a corresponding algorithm and software to deal with the data to reconstruct the wavefunction and display the imagine. This is demanding, especially when we want to realize dynamic object imaging.

The technical challenges noted above, however, will not limit the applications of wavefunction holographic imaging in principle. The challenges are mainly due to technical issues, and there are many experiences we can learn from traditional holographic imaging. Wavefunction holographic imaging provides a totally new approach from the point of view of quantum theory. For living objects that are sensitive to the light intensity that single photons should be used, wavefunction holographic imaging maybe an excellent choice. In addition, there seem to be no technical obstacles for researchers from the area of traditional holographic imaging to access and study this new method based on their existing equipment.

\section{8. Discussion and conclusion}
Currently, all weak value based direct wavefunction reconstructions are limited to the photonic system. The possibility of applying the same method to other physical systems, e.g., electrons or neutrons deserves further consideration. The direct measurement of the wavefunction of electron or neutron would be more physically attractive. It is, however, technically challenging to realize because of the difficulty of spin manipulation in practice. With the development of more advanced technologies in spin manipulation, we believe it is totally possible. 

In conclusion, we have introduced weak value based direct wavefunction reconstruction and discussed its potential application in holographic imaging, which provides an alternative approach beyond traditional ways. The key point of the next stage is to realize the experimental demonstration of 3D objects holographic imaging based on this new method. In addition, wavefunction reconstruction may find important applications in testing foundational issues \cite{superposition} or quantum information science \cite{QC}. As the number of qubits grows quickly in various physical platforms, e.g., superconducting, ionic and atomic quantum processors, the need for efficient quantum state reconstruction seems urgent and the new method introduced may provides a promising way to tackle this issue.

\hfill

\begin{acknowledgments}
\textbf{Acknowledgments:} The authors would like to acknowledge Zhang C R, Xiang G Y, Zhu J, and Wang L W for valuable discussions.
Hu M J is supported by the Beijing Academy of Quantum Information Sciences. Zhang Y S is supported by the
National Natural Science Foundation of China (Grants
No. 11674306 and 92065113) and the University Synergy Innovation Program of Anhui Province (Grants No. GXXT-2022-039).

\end{acknowledgments}


\begin{thebibliography}{0}%
\makeatletter
\providecommand \@ifxundefined [1]{%
 \@ifx{#1\undefined}
}%
\providecommand \@ifnum [1]{%
 \ifnum #1\expandafter \@firstoftwo
 \else \expandafter \@secondoftwo
 \fi
}%
\providecommand \@ifx [1]{%
 \ifx #1\expandafter \@firstoftwo
 \else \expandafter \@secondoftwo
 \fi
}%
\providecommand \natexlab [1]{#1}%
\providecommand \enquote  [1]{``#1''}%
\providecommand \bibnamefont  [1]{#1}%
\providecommand \bibfnamefont [1]{#1}%
\providecommand \citenamefont [1]{#1}%
\providecommand \href@noop [0]{\@secondoftwo}%
\providecommand \href [0]{\begingroup \@sanitize@url \@href}%
\providecommand \@href[1]{\@@startlink{#1}\@@href}%
\providecommand \@@href[1]{\endgroup#1\@@endlink}%
\providecommand \@sanitize@url [0]{\catcode `\\12\catcode `\$12\catcode
  `\&12\catcode `\#12\catcode `\^12\catcode `\_12\catcode `\%12\relax}%
\providecommand \@@startlink[1]{}%
\providecommand \@@endlink[0]{}%
\providecommand \url  [0]{\begingroup\@sanitize@url \@url }%
\providecommand \@url [1]{\endgroup\@href {#1}{\urlprefix }}%
\providecommand \urlprefix  [0]{URL }%
\providecommand \Eprint [0]{\href }%
\providecommand \doibase [0]{http://dx.doi.org/}%
\providecommand \selectlanguage [0]{\@gobble}%
\providecommand \bibinfo  [0]{\@secondoftwo}%
\providecommand \bibfield  [0]{\@secondoftwo}%
\providecommand \translation [1]{[#1]}%
\providecommand \BibitemOpen [0]{}%
\providecommand \bibitemStop [0]{}%
\providecommand \bibitemNoStop [0]{.\EOS\space}%
\providecommand \EOS [0]{\spacefactor3000\relax}%
\providecommand \BibitemShut  [1]{\csname bibitem#1\endcsname}%
\let\auto@bib@innerbib\@empty
\end{thebibliography}%


\begin{thebibliography}{99}
\bibitem{Dirac} Dirac P A M 1958 The Principles of Quantum Mechanics (4nd edn.) (Clarendon, Oxford) pp. 1-18



\bibitem{Beyasian1} Fuchs C A and Schack R 2013  Rev. Mod. Phys. {\bf 85}, 1693

\bibitem{Beyasian2} Pusey M F, Barrett J, and Rudolph T 2012 Nat. Phys. {\bf 8}, 475

\bibitem{Beyasian3} Ringbauer M, Duffus B, Branciard C, Cavalcanti E G, White A G, and Fedrizzi A 2015 Nat. Phys. {\bf 11}, 249

\bibitem{weak} Dressel J, Malik M, Miatto F. M, Jordan A N, and Boyd R W, 2014  Rev. Mod. Phys. {\bf 86}, 307

\bibitem{AAV} Aharonov Y, Albert D Z, and Vaidman L 1988 Phys. Rev. Lett {\bf 60}, 1351

\bibitem{Ludeen} Lundeen J S, Sutherland B, Patel A, Stewart C, and Bamber C 2011 Nature {\bf 474}, 188

\bibitem{QST1} James D F V, Kwiat P G, Munro W J, and White A G 2001 Phys. Rev. A {\bf 64}, 052312

\bibitem{QST2} Lvovsky A I and Raymer M G 2009 Rev. Mod. Phys. {\bf 81}, 299

\bibitem{QST3} Cramer M, Plenio M B, Flammia S T, Somma R, Gross D, Bartlett S D, Landon-Cardinal O, Poulin D, and Liu Y K 2010 Nat. Commun. {\bf 1}, 149

\bibitem{am1} Hosten O and Kwiat P 2008 Science, {\bf 319}, 787-790

\bibitem{am2} Dixon P B, Starling D J, Jordan A N, and Howell J C 2009 Phys. Rev. Lett. {\bf 102}, 173601

\bibitem{am3} Harris J, Boyd R W, and Lundeen J S 2017 Phys. Rev. Lett. {\bf 118}, 070802

\bibitem{Jozsa} Jozsa R 2007 Phys. Rev. A {\bf 76}, 044103

\bibitem{superposition} Hu M J 2022 arXiv:2212.06525[quant-ph] 


\bibitem{strong} Vallone G and Dequal D 2016 Phys. Rev. Lett. {\bf 116}, 040502


\bibitem{direct} Zhang C R, Hu M J, Hou Z B, Tang J F, Zhu J, Xiang G Y, Li C F, Guo G C, and Zhang Y S 2020 Phys. Rev. A {\bf 101}, 012119 

\bibitem{zhang} Zhang C R, Hu M J, Xiang G Y, Zhang Y S, Li C F, and Guo G C 2020 Chin. Phys. Lett. {\bf 37}, 080301

\bibitem{holography1} Zhang T and Yamaguchi I 1998 Opt. Lett. {\bf 23} 1221

\bibitem{holography2} Cuche E, Bevilacqua F, and Depeursinge C 1999 Opt. Lett. {\bf 24}, 291

\bibitem{holography3} Kakue T, Yonesaka R, Tahara T, Awatsuji Y, Nishio K, Ura S, Kubota T, and Matoba O 2011 Opt. Lett. {\bf 36}, 4131

\bibitem{POVM} Kraus K 1983 States, Effects, and Operations (Springer-Verlag, Berlin) pp. 1-30



\bibitem{free} Shi Z M, Mirhosseini M, Margiewicz J, Malik M, Rivera R, Zhu Z Y, and Boyd R W 2015 Optica {\bf 2}, 4 




\bibitem{phase} Javidi B, Carnicer A, Anand A{\it et al.} 2021 Opt. Express {\bf 29}, 35078


\bibitem{BPM} Kamilov U S, Papadopoulos I N, Shoreh M H, Goy A, Vonesch C, Unser M, and Psaltis D 2016 IEEE Transactions on Comput. Imaging {\bf 2}, 59

\bibitem{QC} Zhou Y, Zhao J, Hay D, McGonagle K, Boyd R W, and Shi Z 2021 Phys. Rev. Lett. {\bf 127}, 040402



\end{thebibliography}
\end{document}